\documentclass[mathleft
]{an}
\usepackage{graphicx}
\usepackage{times}
\begin{document}

% The following seven commands are intended for editorial usage and should be ignored by
% the author(s).
\Pagespan{789}{}% Document's page range. 
% If second parameter is left empty, the last page is computed automatically.
\Yearpublication{2006}%
\Yearsubmission{2005}%
\Month{11}%   
\Volume{999}%  
\Issue{88}% 
% \DOI{This.is/not.aDOI}% 

\title{High-resolution X-ray spectroscopy: the coming-of-age}

\author{J.S. Kaastra\inst{1}\fnmsep\thanks{Corresponding author:
  \email{j.s.kaastra@sron.nl}\newline}
%Example 
%for footnote, note the usage of the \texttt{fnmsep}
%command as separator between institute number and footnote mark} 
%\and  G.H. Ostwriter\inst{2,3}
}
\titlerunning{High-resolution X-ray spectroscopy}
\authorrunning{J.S. Kaastra}
\institute{
SRON Netherlands Institute for Space Research, Sorbonnelaan 2,
           3584 CA Utrecht, the Netherlands 
           \and
           Leiden Observatory, Leiden University, PO Box 9513, 
           2300 RA Leiden, the Netherlands
           \and
           Department of Physics and Astronomy, Universiteit Utrecht, 
           P.O. Box 80000, 3508 TA Utrecht, the Netherlands
}

\received{\today}
\accepted{TBD}
\publonline{later}


\abstract{ 
Since the launch of Chandra and XMM-Newton, high-resolution X-ray spectra of
cosmic sources of all kinds have become available. These spectra have resulted
in major scientific breakthroughs. However, due to the techniques used, in
general high-quality spectra can only be obtained for the brightest few sources
of each class. Moreover, except for the most compact extended sources, like cool
core clusters, grating spectra are limited to point sources. Hitomi made another
major step forward, in yielding for the first time a high-quality spectrum of an
extended source, and improved spectral sensitivity in the Fe-K band. For point
sources with the proposed Arcus mission, and for all sources with the launch of
Athena, X-ray spectroscopy will become mature. It allows us to extend the
investigations from the few handful of brightest sources of each category to a
large number of sources far away in space and time, or to get high
time-resolution, high-spectral resolution spectra of bright time variable
sources.
}

\maketitle

\section{Introduction}
XMM-Newton has been operational already for more than 16 years. It has made
many major achievements with all of its instruments, and as we show in this
paper, it is still capable of making great discoveries for the next decade.
In this contribution we focus on high-resolution spectroscopy. We briefly
mention some spectroscopic highlights of XMM-Newton. We will show the great
potential offered by the Hitomi satellite during its short lifetime, and then
focus on the prospects for some future X-ray missions. We conclude with
prospects for RGS for the next decade.

\section{X-ray instruments}
Present-day X-ray instruments can be subdivided into two classes. The most
common instruments have relatively high throughput (effective area) but medium
to low spectral resolution. In addition, they can do imaging. Good examples of
these instruments are the CCD cameras on XMM-Newton, Chandra, Suzaku and Swift.
They constitute the most commonly used type of instrument, because of their
sensitivity and imaging capabilities. Best use of them is made in studies of
broad-band variability, imaging and surveys of large areas or large number of
sources.

For detailed astrophysical modeling, often higher spectral resolution is needed.
This is until recently offered exclusively by grating spectrometers, such as the
RGS spectrometer on XMM-Newton, and the LETGS and HETGS spectrometers on board
of Chandra. Compared to CCDs, their imaging capabilities are limited (mostly in
the cross-dispersion direction of the grating), and their effective area lower.
They are most usefull for the brightest sources of each class. However, provided
sufficient source flux, the amount of astrophysical information that can be
obtained from a grating spectrum is very large. Due to the way gratings are
designed, they are perfect for point sources but show degraded resolution for
spatially extended sources. RGS is currently the best grating spectrometer for
extended sources, but it can cope only with sources that are effectively smaller
than a few arcminutes, like cool core clusters.

Before considering the prospects for XMM-Newton in the future, we first present
a few examples of what has been achieved now with this mission with some of the
most prestiguous projects.

\section{XMM-Newton now}

XMM-Newton has an impressive track record of scientific discoveries. We have no
room to discuss that here in detail. We only mention a few highlights here, and
only from the RGS. The most important discovery was the discovery of the lack of
large amounts of cooling gas in cool core clusters (Peterson et al. 2001; Tamura
et al. 2001; Kaastra et al. 2001). Other important early work was the discovery
of soft X-ray line emission from photoionised gas in Seyfert 2 galaxies
(Kinkhabwala et al. 2002), and the discovery of unresoved inner shell lines from
lowly ionised iron in AGN outflows (Sako et al. 2001), not to mention the
important work on stellar coronae (e.g. Brinkman et al. 2001) and other sources.
All these papers have $\sim$150 citations or more.

XMM-Newton in general and RGS in particular remain capable of making new
breakthroughs. The same holds for the grating spectrometers on Chandra. We only
mention here the following recent highlights of grating spectroscopy, showing
the power of these 17 year old missions: the unambiguous detection of the WHIM
in absorption (Nicastro et al. 2013), outflows from a tidal disruption event
(Miller et al. 2015) and the discovery of ultra-fast outlfows from
Ultra-Luminous X-ray sources (Pinto et al. 2016).

Other prominent work that becomes increasingly more important now the mission
ages are the use of large data sets or deep exposures of individual targets, and
this is the obvious way to make further progress. 

A good example is AGN monitoring, where multiple observations with proper
spacing allow to use the physical information contained in the time variability
to derive for instance the location of the different spectral components (e.g.
Kaastra et al. 2012). The associated stacked spectra yield full insight into the
landscape of the AGN environment (Detmers et al. 2011).

Another type of study is the study of the chemical composition of clusters of
galaxies; stacking the data of a large number of clusters, and using both RGS
and EPIC data, Mernier et al. (2016ab) was able to obtain the most accurate
decomposition so far.

Stacked cluster data also can be used for a search of dark matter (Bulbul et al.
2014). This study, focused on the possible detection of an unidentified 3.5~keV
line that might be attributed to the decay of sterile neutrinos, a possible form
of dark matter, also triggered work on charge exchange emission in clusters (Gu
et al. 2015), a more common explanation for the presence of a 3.5~keV line. All
this has been possible thanks to the archive of 16 years of XMM-Newton
observations.

\section{Hitomi}
With the launch of Hitomi on February 17 from Japan, a new perspective for
high-resolution X-ray spectroscopy has opened. The three strongest features of
this mission are the following.

First, it is now possible to obtain high-resolution spectra also for spatially
extended sources, thanks to the SXS detector, a 6x6 pixel calorimeter array. The
first light for this instrument was the Perseus cluster (Hitomi collaboration
2016). For the first time it was possible to directly determine the turbulence
in a cluster of galaxies by measuring the line broadening. The measured
turbulent velocity in Perseus was 164$\pm 10$~km/s. Before this, only rather
crude upper limits of order 500~km/s were obtained using RGS spectra of clusters
(Pinto et al. 2015) or indirect measurements using resonance scattering with RGS
could constrain the turbulence somewhat (Werner et al. 2009, De Plaa et al.
2012). However, in the latter case not fully resolved issues with the precise
atomic data made the analysis difficult, and in the former case the relatively
low effective spectral resolution for extended sources and instrumental line
spread function uncertainties were limiting the conclusions.

The second strength of the SXS instrument of Hitomi was the high spectral
resultion in the Fe-K band, not only for clusters of galaxies like Perseus, but
also for point sources like active galactic nuclei or X-ray binaries. A barely
detected (2$\sigma$) narrow Fe-aborption line in the prototype Seyfert 1 outflow
galaxy NGC~3783 using a Ms of Chandra HETGS data would have been surpassed by
SXS: in only 20\% of that exposure time, it would have rendered a 20$\sigma$
detection. Unfortunately, the Hitomi satellite failed before the first AGN
spectrum could be taken.

A third strength, namely simultaneous broad-band sensitivity over the band
between 0.3 keV and several 100 keV could also not been proven due to the
unfortunate fate of the mission.

Already the Hitomi spectrum of the Perseus cluster has challenged the atomic
plasma codes that are being used widely in the X-ray community. Several updates
in both the SPEX and APEC codes were needed to get the last bits of information
out of this spectrum, and at this spectral resolution, even things like Voigt
line profiles, rather than Gaussian approximations, start becoming relevant.

\section{Future X-ray spectroscopy missions} 
Due to the unfortunate end of Hitomi, at present only the almost 17 year old
grating spectrometers of XMM-Newton and Chandra are our spectroscopic working
horses. They are still doing fine, yielding regularly new discoveries (Sect.~3),
but we are waiting for other breakthroughs facilitated by new missions.

A Hitomi recovery mission, if approved, could fly early in the next decade
around 2020/2021 and would do all the science foreseen for the SXS instrument.
Other ideas, for wide field of view, high spectral resolution missions are being
considered, like DIOS in Japan or comparable missions in China, but in the
former case this may not be independent of the selection of a Hitomi successor.
Time will show.

\begin{table}[!tbp]
\caption{Resolving power ($R$) and figure of merit $F\equiv \sqrt{RA_{\rm eff}}$
for weak line detection for existing and future missions compared to Hitomi
(which has spectral resolution 5~eV for all energies).}
\label{tab:rfom}
\begin{tabular}{lcccccc}
Instrument & $R$ & $R$ & $R$ & $F$ & $F$ & $F$ \\
E (keV) & 0.5 & 1.5 & 6 & 0.5 & 1.5 & 6 \\
\hline
HRC-S/LETG  & 6 & 0.7 & 0.04 & 1.3 & 0.2 & 0.03 \\
ACIS-S/LETG & 6 & 0.7 & 0.04 & 0.5 & 0.4 & 0.08 \\
MEG         & 12 & 1.4 & 0.09 & 0.3 & 0.6 & 0.06 \\
HEG         & -- & 3 & 0.17 & -- & 0.5 & 0.13 \\
RGS         & 4 & 0.5 & -- & 3 & 0.24 & -- \\
SXS Hitomi  & $\equiv 1$ & $\equiv 1$ & $\equiv 1$ & 
              $\equiv 1$ & $\equiv 1$ & $\equiv 1$\\
Arcus       & 25 & 8 & -- & 17 & 3.5 & -- \\
Athena XIFU & 2 & 2 & 2 & 19 & 9 & 5 \\
\hline
\end{tabular}
\end{table}

ESA is preparing the Athena mission, a major new flagship to be launched in
2028, with superb capabilities in terms of effective area, imaging qualities and
spectral resolution as compared to XMM-Newton. That promising mission is
discussed elsewhere (Nandra et al. 2013). Here I want to focus on another
mission that might be launched earlier, Arcus. Table~\ref{tab:rfom} compares
the relative strengths of several X-ray missions in terms of resolving power and
figure of merit for weak line detection.

\section{The Arcus mission}

Arcus is a concept for a NASA Midex proposal; if selected it could fly around
2023. A more detailed description of this mission can be found in Smith et al.
(2016). Basically, one may consider it in terms of performance as a mission with
ten times the spectral resolution and ten times the effective area of the RGS,
operating over a slightly different band between roughly 10--50~\AA. 

The main science goals are a study of the halo of our Milky way and other
galaxies by measuring their absorption spectra towards bright background
sources, AGN feedback, and several other interesting science questions. Here I
focus on AGN feedback.

\begin{figure}[!tbp]
\resizebox{0.9\hsize}{!}{\includegraphics[angle=-90]{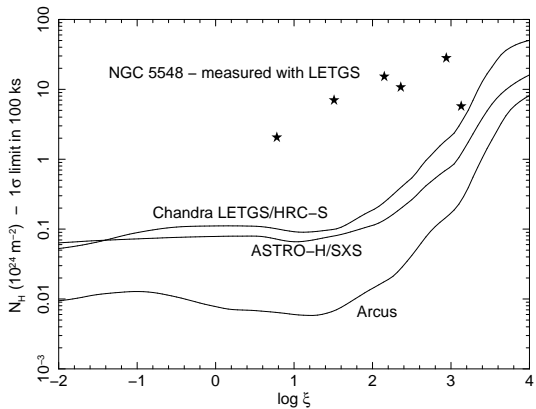}}
\caption{Minimum detectable column density of an outflow in 100~ks at the
1$\sigma$ level. See text for details.
}
\label{fig:collim}
\end{figure}

Due to the increased sensitivity of Arcus compared to the Chandra or XMM-Newton
gratings, much lower column densities can be detected, as illustrated in
Fig.~\ref{fig:collim}. We have considered here the column densities of the six
photoionised components of the outflow in NGC~5548 in its normal, unobscured
state (Kaastra et al. 2014). These are indicated by stars in the figure.  We
have simulated this spectrum for each of the following instruments for 100~ks
exposure time: Chandra LETGS, ASTRO-H (Hitomi) SXS detector, and Arcus. These
simulated spectra were fitted over the full band of the instruments with the
same model, and the nominal statistical uncertainty of the column density has
been determined and is shown in Fig.~\ref{fig:collim}. For ionisation parameters
$\log \xi < 2$ Arcus is about an order of magnitude more sensitive than the
other instruments, while for the higher ionisation parameters it is still more
sensitive but to a lesser extend.

\begin{figure}[!tbp]
\resizebox{0.9\hsize}{!}{\includegraphics[angle=-90]{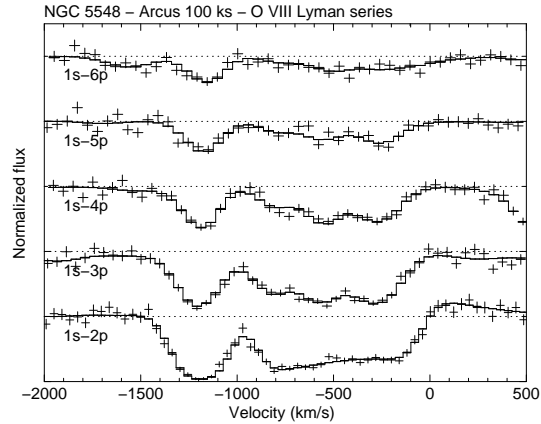}}
\caption{Normalized spectra (continuum set to unity) for a 100~ks exposure of
NGC~5548 in its normal unobscured state. Individual spectra have been shifted
for clarity. The Lyman series of O~VIII is shown.}
\label{fig:o8}
\end{figure}

We show the Arcus capabilities for the same source in another way in
Fig.~\ref{fig:o8}. It is easy to detect the full Lyman series of ions such as
O~VIII, and to resolve the lines into all their velocity components. This is a
new aspect of Arcus: with the existing grating data (Chandra and XMM-Newton),
these lines are not resolved; they only show a small amount of broadening, and
from the centroid shifts one may get some inpression whether the components at
$-1200$~km\,s$^{-1}$ are more dominant than those around $-500$~km\,s$^{-1}$.
With Arcus we will resolve the spectra down to about 100~km\,s$^{-1}$, close to
their width as can be determined e.g. from UV spectra. However, the great
advantage of X-ray spectra compared to UV spectra is the much larger number of
ions that are available for diagnostic purposes; in the UV band only a small
number of low-ionisation lines can be investigated. Moreover, feedback is
stronger at the high ionisation X-ray lines, both due to higher column densities
and higher velocities (e.g. Ultra-Fast Outflows).

\begin{figure}[!tbp]
\resizebox{0.9\hsize}{!}{\includegraphics[angle=-90]{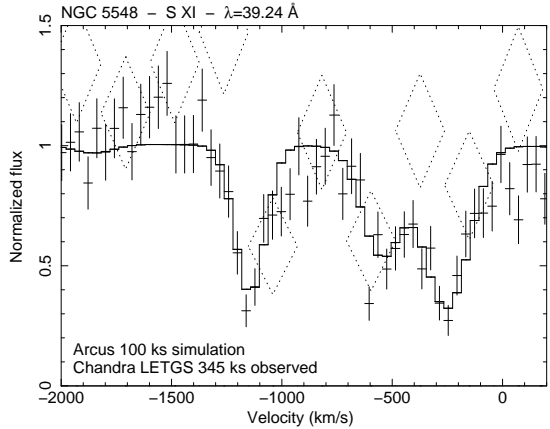}}
\caption{Observed Chandra LETGS spectrum of NGC~5548 in its normal, unobscured
state near the S~XI line (exposure time 345~ks, dotted diamonds) and a 100~ks
simulation of the same spectrum with Arcus (crosses). The model used for the
simulation is indicated by the solid line and is based on the Chandra data (see
Kaastra et al. 2014 for details). 
}
\label{fig:s11}
\end{figure}

Fig.~\ref{fig:s11} shows a part of the Chandra LETGS spectrum of NGC~5548 in its
normal state. This instrument is the only instrument so far that is capable of
measuring the spectrum beyond 38~\AA. It is focused on the strongest S~XI line
near 39.24~\AA. This line was first identified in this spectrum by Steenbrugge
et al. (2005), but given the relatively low effective area of the LETGS the
detection is not very strong, and rests upon the prediction that it should be
there based on stronger lines from other ions at shorter wavelengths. As the
figure shows, Arcus will have no problems in detecting the line and even in
resolving it into its velocity components that span the range from $-1200$ to
0~km\,s$^{-1}$. This line is of interest, because it has nearby lines from the
same ion that arise from metastable levels, hence S~XI is an excellent density
indicator. We elaborate that later in this section.

\begin{figure}[!tbp]
\resizebox{0.9\hsize}{!}{\includegraphics[angle=-90]{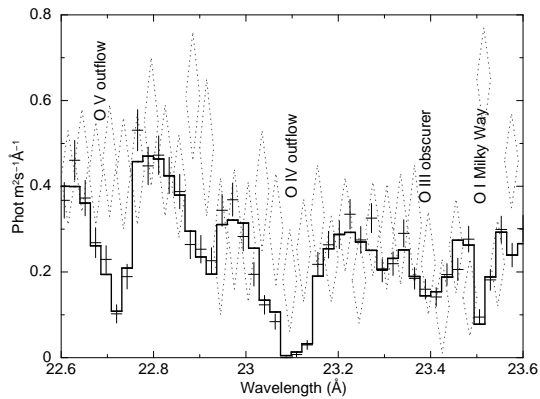}}
\caption{Part of a simulation of a 660 ks Arcus observation of NGC~5548 in its
obscured state (solid line: model; solid crosses: simulated data), compared to
the observed 660~ks RGS spectrum (dotted diamonds). The Arcus data have been
rebinned by a factor of 10 for this plot in order to match better the RGS
resolution. Two absorption lines from the outflow (O~V and O~IV), a line from
the obscurer (O~III) and a Galactic foreground line from O~I are indicated.}
\label{fig:obscurer}
\end{figure}

For decades NGC~5548 has been in a state showing a prototypical outflow that was
used for the simulations shown in Figs.~\ref{fig:o8}--\ref{fig:s11}. However,
starting around 2011, the source showed in addition strong obscuration by
high-column density, low ionisation material (Kaastra et al. 2014). This
obscuring material was close to the broad line region, contrary to the normal
outflow that is located at pc-scale distances (Arav et al. 2015, Ebrero et al.
2016). 

The obscurer caused a strong lowering of the ionisation parameter of this normal
outflow. The direct observational evidence for the obscuring material comes from
the UV band, showing broad (5000~km\,s$^{-1}$) outflowing absorption in
intermediate ionisation lines, and in X-rays mainly from the continuum opacity
of the obscuring material. Because the obscuration is not 100\%, a small amount
of X-rays still leaks through the soft X-ray band, but the remaining flux is
roughly 20 times less than the ''normal'', unobscured flux in this band. Due to
this effect, RGS was unable to reveal the details of the obscured soft X-ray
absorption spectrum; only the emission lines from a region far away from the
nucleus remained unobscured and visible. However, the best-fit model predicts
that in the soft X-ray band both absorption lines from the normal outflow (at a
reduced ionisation parameter) and from the obscurer should be present. This is
shown in Fig.~\ref{fig:obscurer}. The figure shows how well these lines can be
seen, and their profiles are resolved by Arcus. This cannot be done with
Athena (too low spectral resolution).

\begin{figure}[!tbp]
\resizebox{0.9\hsize}{!}{\includegraphics[angle=-90]{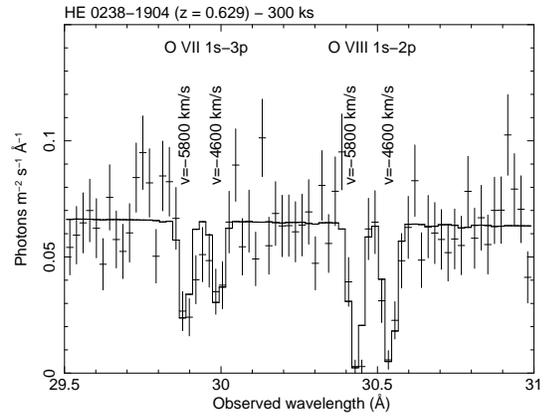}}
\caption{Arcus simulation of the spectrum of HE~0238-1904.}
\label{fig:he}
\end{figure}

The enhanced sensitivity of Arcus not only allows for better (or time-resolved)
spectra of the brightest sources, it also allows to study outflows at much
larger distances. This is illustrated in Fig.~\ref{fig:he}, where we show a
300~ks simulation of a redshift 0.629 quasar near the oxygen K-complex. Due to
the relatively high redshift, these lines are redshifted from the 19~\AA\ band
to the 30~\AA\ band. The simulation is based on the UV spectrum published by
Arav et al. (2013). Velocities and ionisation stages can be well separated.

\begin{figure}[!tbp]
\resizebox{\hsize}{!}{\includegraphics[angle=-90]{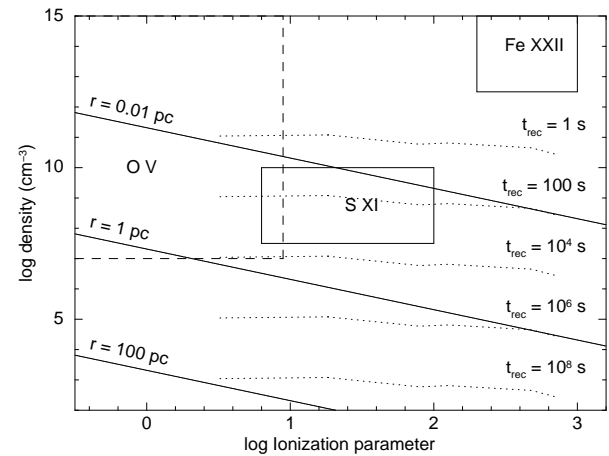}}
\caption{Density indicators for AGN outflows. See text for details.}
\label{fig:dens}
\end{figure}

Finally, the enhanced sensitivity of Arcus and its relatively broad wavelength
range also will open the doors for a completely new type of diagnostic in AGN
outflows: using density sensitive lines. Up to now, density sensitivity has been
seen only in some grating spectra of very bright X-ray binaries, e.g. Miller et
al. (2008).

In Fig.~\ref{fig:dens} we give a few examples of density dignostics for Arcus
absorption spectra of AGN, again based on the normal, unobscured spectrum of
NGC~5548. In the ionisation parameter versus density plane we plot the regions
for which we have density diagnostics. Solid lines indicate lines of equal
distance to the ionising source. The dotted lines indicate the densities that
can be derived from variability studies: by constraining the recombination time
scale through measuring or limiting the lag of the outflow spectrum relative to
continuum flux variations. See e.g. Ebrero et al. (2016) for more details. This
technique can also be used with current instruments, however it requires a
substantial amount of monitoring observations. New are the density diagnostics
by O~V K-shell lines (Kaastra et al. 2004), indicated by the dashed region, and
in particular S~XI and Fe~XXII. S~XI has a few lines in the 39~\AA\ part of the
spectrum that are sensitive to relatively low densities; Fe~XXII is sensitive to
much higher densities.

\section{What can we do now?}

The plans for future missions like a successor for Hitomi, Arcus or Athena are
great. Still, at least for the first four years we will have no new spectroscopy
mission. Some of the science that we outlined earlier really needs the
high-resolution and high throughput, but long, deep spectra of bright sources
taken with XMM-Newton can already give new insights. This holds for most classes
of objects, as long as systematic calibration limits are not reached. We come
back to that in Section~\ref{sect:systematic}.

For these deep exposures, all three main instruments of XMM-Newton (RGS, EPIC
and the OM) are of importance. A good example of the latter is formed by deep
monitoring campaigns of active galactic nuclei. These sources vary on a
multitude of time scales: from minutes to decades. In most cases the archive
only contains poorly sampled data, or data sampled only at one particular time
scale. However, monitoring campaigns cover such time scales and give excellent
time-averaged spectra (for Ms-scale exposure times).

The major strength of the XMM-Newton instruments in such campaigns is for EPIC
its sensitivity to variability as well as its broad energy range, for RGS the
high spectral quality and for OM the additional UV spectrum that is essential to
know for photoionisation modelling.

Such campaigns are even more important when they are performed as part of a
multi-wavelength campaign including other instruments. For AGN outflow studies,
in particular HST with its COS spectrograph, NuSTAR with its high-energy
coverage, Swift with its snapshot capability and ground-based facilities are the
most important components, apart of course from the XMM-Newton data that form
the heart of such campaigns.

With a predicted lifetime of another decade, XMM-Newton has good prospects for
performing a number of such projects. However, it is not well known how long the
other facilities needed for such work will be operational. Therefore caution
must be taken not to delay such projects to near the real end of the mission.

\section{Limitations imposed by systematic effects\label{sect:systematic}}

\begin{figure}[!tbp]
\resizebox{\hsize}{!}{\includegraphics[angle=-90]{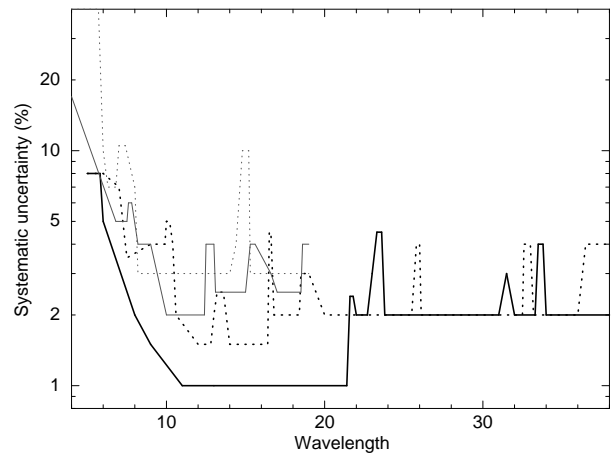}}
\caption{Ultimate systematic uncertainties of the RGS effective area as a
function of wavelength. Solid lines: RGS1; dashed lines: RGS2; thick lines:
first order spectra; thin lines: second order spectra.}
\label{fig:sys}
\end{figure}

Although still a lot can be learned by making deeper exposures of X-ray sources,
there is a natural limitation to how deep one can go. That is given by the
exposure time where the statistical uncertainties on the spectrum become smaller
than the systematic uncertainties. We have investigated this for RGS as follows.

Within the framework of the RGS calibration, we have derived time- and
wavelength dependent correction factors for the RGS effective area. These
corrections are based on a detailed study of about a hundred RGS spectra of the
blazars Mrk~421 and PKS~2155-304. Space does not allow us to discuss that in
detail here. After applying these effective area corrections, there still
remains some scatter in the fit residuals for individual spectra. A part of this
is of statistical nature and hence of no real concern. However, after correcting
for the statistical uncertainties the remaining scatter can be attributed to
ultimate systematic uncertainties. We show these uncertainties in
Fig.~\ref{fig:sys}. 

It is seen that the first order RGS1 spectra have systematical uncertainties of
1--2\% over most of the range, while RGS2 has typically 2\%. Exceptions are
found in some narrower wavelength ranges that are likely due to some remaining
slightly unstable pixels or near 32~\AA\ due to time-variable nitrogen
contamination. Other exceptions are found at the shortest wavelengths, where it
is difficult to model the broad-band grating scattering accurately. For more
details about RGS calibration, see also De Vries et al. (2015).

\begin{figure}[!tbp]
\resizebox{\hsize}{!}{\includegraphics[angle=-90]{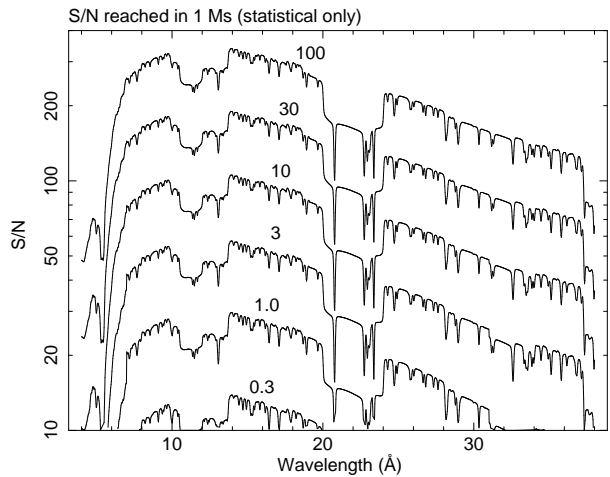}}
\caption{Signal to noise ratio per 0.06~\AA\ resolution element for the combined
first order RGS detectors that can be achieved in 1~Ms for a continuum source
with a flux as labeled. Units of the flux are
photons\,m$^{-2}$\,s$^{-1}$\,\AA$^{-1}$.}
\label{fig:sn}
\end{figure}

As a next step, we have determined the statistical uncertainties of the flux
within a single resolution element of 0.06~\AA\ width, that can be
achieved for a source of a given brightness (Fig.~\ref{fig:sn}). In these
calculations, a typical quiescent background has been taken into account. That
causes the relative low values for the lowest flux considered here of
0.3~photons\,m$^{-2}$\,s$^{-1}$\,\AA$^{-1}$. For even lower fluxes the
background becomes dominant. Obviously, it does not make sense to observe the
brightest example here (100~photons\,m$^{-2}$\,s$^{-1}$\,\AA$^{-1}$) for a Ms,
because the peak S/N of about 300 corresponds to an uncertainty of 0.3\%, much
smaller than the systematic uncertainty of the effective area.

\begin{figure}[!tbp]
\resizebox{\hsize}{!}{\includegraphics[angle=-90]{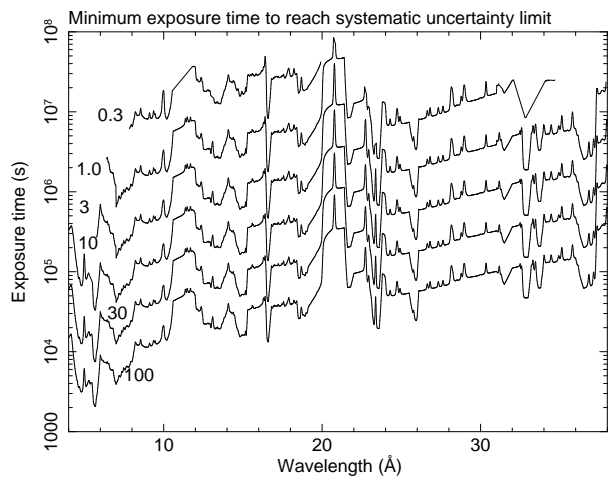}}
\caption{Minimum exposure time needed to get a statistical uncertainty per
0.06~\AA\ resolution element equal to the systematic uncertainty in the RGS
effective area, for the combined first order RGS detectors, for sources  with a
flux as labeled. Units of the flux are photons\,m$^{-2}$\,s$^{-1}$\,\AA$^{-1}$.}
\label{fig:minexp}
\end{figure}

For that reason, we have calculated for sources with different fluxes the
required exposure time to reach this systematic limit (Fig.~\ref{fig:minexp}).
For the brightest source this is between 10 and 100~ks; for a typical bright AGN
(10~photons\,m$^{-2}$\,s$^{-1}$\,\AA$^{-1}$) exposure times of a few 100~ks up
to about a Ms make sense. Obviously, one can always observe these sources longer
if variability rather than spectral quality is the main science driver.

\section{Conclusions}

More than 16 years of XMM-Newton spectroscopy has delivered fascinating science.
Even now new topics appear, triggered by carefully investigating the large
available databases, serendipitous discoveries and new views made possible by
other facilities. While waiting for new missions, XMM-Newton can make
significant progress by going deeper and longer in the coming decade.

\acknowledgements

SRON is supported financially by NWO, the Netherlands Organization for
Scientific Research.   

%\newpage%%%%%%%%%%%%%%%%%%%%%%%%%%%%%%%%%%%%%%%%%%%%%%%%%%%%%%

\end{document}